\documentclass[amsmath,amssymb,12pt]{article}
\usepackage{jheppub}
\pdfoutput=1
\usepackage[utf8]{inputenc}
\usepackage{amsfonts}
\usepackage{graphicx}
\usepackage{slashed}
\usepackage{subfig}
\usepackage{array}

\usepackage{color}

\newcommand{\Ht}{\mathbb{H}^{3}}

\newcommand*{\affmark}[1][*]{\textsuperscript{#1}}

\newcommand{\red}[1]{{\color{red}{#1}}}
\newcommand{\blue}[1]{{\color{blue}{#1}}}
\newcommand{\mage}[1]{{\color{magenta}{ #1}}}

\newcommand{\beq}{\begin{equation}}

\newcommand{\eeq}{\end{equation}}

\begin{document}

\title{BTZ one-loop determinants via the Selberg zeta function for general spin}

\author{Cynthia Keeler\affmark[1], Victoria L. Martin\affmark[1] and Andrew Svesko\affmark[1]}
\affiliation{\affmark[1]Department of Physics, Arizona State University, 
Tempe, Arizona 85287, USA}
\emailAdd{keelerc@asu.edu}
\emailAdd{victoria.martin.2@asu.edu}
\emailAdd{asvesko@asu.edu}

\abstract{We relate the heat kernel and quasinormal mode methods of computing the 1-loop partition function of arbitrary spin fields on a rotating (Euclidean) BTZ background using the Selberg zeta function associated with $\Ht/\mathbb{Z}$, extending (1811.08433) \cite{Keeler:2018lza}. Previously, Perry and Williams \cite{Perry03-1} showed for a scalar field that the zeros of the Selberg zeta function coincide with the poles of the associated scattering operator upon a relabeling of integers. We extend the integer relabeling to the case of general spin, and discuss its relationship to the removal of non-square-integrable Euclidean zero modes.} 


\maketitle

\section{Introduction}
\label{intro}

Constructing a complete theory of Euclidean quantum gravity requires knowledge of the full partition function
\beq\label{Z}
Z=\int DgD\phi~ e^{-S_{E}(g,\phi)/\hbar}\;,
\eeq
where $g$ is the dynamical metric and $\phi$ represents all other matter fields. Although the partition function $Z$ is often intractable to compute directly, it can be evaluated perturbatively by expanding the Euclidean action $S_{E}$ about a classical solution using a saddle point approximation. This approximation is an asymptotic expansion in $\hbar$, where $Z^{(0)}\sim\mathcal{O}(\hbar^{-1})$ is the classical contribution to the full partition function $Z$, and $Z^{(1)}\sim\mathcal{O}(\hbar^{0})$ captures leading order 1-loop quantum effects. For a free field $\phi$ on a gravitational background $\mathcal{M}$, computing the 1-loop partition function $Z_{\phi}^{(1)}$ involves calculating functional determinants of kinetic operators $\nabla_{\phi,\mathcal{M}}^{2}$. For example, when $\phi$ is a complex scalar field,
\beq 
Z_{\phi}^{(1)}=[\text{det}(-\nabla_{\phi,\mathcal{M}}^{2})]^{-1}\;.
\label{1loopdetbasic}
\eeq
This perturbative approach has proven useful in finding quantum corrections to black hole entropy \cite{Sen:2008vm,Banerjee:2010qc,Sen:2011ba} and holographic entanglement entropy \cite{Barrella:2013wja}. Functional determinants are also of pure mathematical interest as their spectral properties provide a classification of smooth manifolds \cite{Nakahara:2003nw}.

There are several methods for computing functional determinants of kinetic operators such as those found in (\ref{1loopdetbasic}). In this work we focus on further developing a connection between two particular methods, referred to in the literature as the heat kernel method (c.f. \cite{Giombi:2008vd}) and the quasinormal mode method \cite{Denef:2009kn}. We reserve a review of these methods for Section \ref{reviewstuff}.

Recently we showed \cite{Keeler:2018lza} how to connect the heat kernel and quasinormal mode methods for computing 1-loop determinants of scalar, vector and tensor fields on locally thermal $\text{AdS}_3$ spacetimes, namely the Ba\~nados, Teitelboim and Zanelli (BTZ) black hole \cite{Banados:1992wn} and thermal $\text{AdS}_3$. In particular, \cite{Keeler:2018lza} showed that the two methods can be formally related via the Selberg zeta function \cite{Patterson89-1} --- a zeta function that is built entirely from the quotient structure of $\mathbb{H}^{3}/\mathbb{Z}$:
\beq
Z_{\mathbb{Z}}(z)=\prod_{k_{1},k_{2}=0}^{\infty}\left[1-e^{2ibk_{1}}e^{-2ibk_{2}}e^{-2a(k_{1}+k_{2}+z)}\right]\;.
\label{selbpatbasic}
\eeq
Here $a$ and $b$ correspond to parameters of the group generator\footnote{For example, in the case of the rotating BTZ black hole $a$ and $b$ are related to the inner and outer radii of the solution, $r_+$ and $r_-$. Namely, $a=\pi r_+$ and $b=\pi|r_-|$.} $\gamma\in\mathbb{Z}$. Specifically, for a scalar field on a BTZ background, the 1-loop partition function $Z^{(1)}(\Delta)$ can be recast in terms of the Selberg zeta function $Z_{\mathbb{Z}}(\Delta)$ via the heat kernel method. We then showed that identifying the zeros of the Selberg zeta function $z^{\ast}$ with the conformal dimension $\Delta$ of the field in question is equivalent to requiring the quasinormal mode frequencies $\omega_{\ast}$ be equal to the Matsubara frequencies $\omega_{n}(T)$:
\beq
 \Delta=z^{\ast}\Longleftrightarrow\omega_{\ast}(\Delta)=\omega_{n}(T)\;.
\label{equivcond}\eeq
We also showed a similar relationship for spin-1 and spin-2 fields. The condition (\ref{equivcond}) leads to the following observation: if any two of (i) Selberg zeta function, (ii) Matsubara frequencies, or (iii) quasinormal modes of the given spacetime are known, the third can be reconstructed. This observation provides a means of predicting quasinormal modes, or Matsubara frequencies of fields on locally thermal $\text{AdS}_{3}$ spacetimes.

In this article we generalize the results of \cite{Keeler:2018lza} to include arbitrary spin fields. Specifically, we express the 1-loop partition function for an arbitrary spin-$s$ field in terms of Selberg zeta functions and show how setting the arguments to the zeros of the Selberg zeta function lead to $\omega_{n}(T)=\omega_{\ast}(\Delta_{s})$. 
Previously it was shown that, for the case of a scalar field and upon appropriate relabeling of integers $k_{1}, k_{2}$ in (\ref{selbpatbasic}), the zeros of the Selberg zeta function coincide with the poles of the associated ``scattering" operator on $\Ht/\mathbb{Z}$ \cite{Perry03-1}. We propose a generalization of this relabelling for general spin, and show that the form of our relabeling is a consequence of removing non-square-integrable Euclidean zero modes from the set of possible Euclidean solutions to field equations.

The article is organized as follows. A brief technical review of the heat kernel and quasinormal mode methods 
for arbitrary spin fields on $\mathbb{H}^{3}/\mathbb{Z}$ is given in Section \ref{reviewstuff}. Section \ref{sec:connectingonH3} connects the heat kernel and quasinormal mode methods for arbitrary spin fields on $\Ht/\mathbb{Z}$ via the Selberg zeta function. 
Concluding remarks and an outline for future work is given in Section \ref{disc}. 


\section{Review: Heat Kernels and Quasinormal Modes on $\mathbb{H}^{3}/\mathbb{Z}$}
\label{reviewstuff}

\subsection{Heat Kernel Method}

In calculating functional determinants det(-$\nabla^{2}_{\psi, \mathcal{M}}$), we seek solutions to  
\begin{equation}
\nabla^{2}_{\mathcal{M}}\psi_{n}=E_{n}\psi_{n}.
\end{equation}
In this review section we employ a kinetic operator with a discrete spectrum simply to demonstrate the heat kernel method. The method also applies for non-compact manifolds, where the spectrum of $\nabla^{2}_{\mathcal{M}}$ is continuous. In this case a divergent contribution proportional to the volume of $\mathcal{M}$ appears. Such divergences are generally removed by subtracting a ``reference'' heat kernel \cite{Vassilevich:2003xt}. The heat kernel method \cite{Giombi:2008vd,David:2009xg} involves constructing an object $K^{(s)}_{ab}(x;y;t)$ dependent on two spacetime points $x$ and $y$ on $\mathcal{M}$: 
\beq K^{(s)}_{ab}(x;y;t)\equiv\langle y,b|e^{-t\nabla^{2}_{(s)}}|x,a\rangle=\sum_{n}\psi^{(s)}_{n,a}(x)\psi^{(s)\ast}_{n,b}(y)e^{-tE_{n}^{(s)}}\;.\label{hkarbspinbasic}\eeq
Subscripts $a$ and $b$ denote indices of the ($2s+1$)-dimensional representation of $SL(2,\mathbb{C})$ of the spin-$s$ field \cite{David:2009xg}. By construction, $K^{(s)}_{ab}(x,y;t)$ satisfies the heat equation
\begin{equation}
(\partial_{t}+\nabla^{2}_{x})K^{(s)}_{ab}(x,y;t)=0, \qquad K^{(s)}_{ab}(x,y;0)=\delta(x,y)\;. \label{heateqn}\end{equation} 
In fact, the eigenfunction expansion in (\ref{hkarbspinbasic}), or solving the heat equation (\ref{heateqn}), may be considered to be two different methods of constructing the heat kernel, though they coincide. Below we will outline the eigenfunction expansion approach.

Knowing the heat kernel allows us to calculate the 1-loop partition function $Z^{(1)}_{s}$ 
\beq \log Z^{(1)}_{s}=\log\text{det}(-\nabla^{2}_{(s)})=-\int^{\infty}_{0}\frac{dt}{t}K^{(s)}(t)\;,\eeq
where
\beq K^{(s)}(t)=\text{tr}(e^{-t\nabla^{2}_{(s)}})=\int_{\mathcal{M}}\sqrt{g}~d^{d+1}x\sum_{a}K_{aa}^{(s)}(x,x;t)\;\label{hkastracegens}\eeq
is called the coincident heat kernel.

Obtaining the heat kernel on generic $\mathcal{M}$ is difficult. However, for highly symmetric backgrounds such as hyperbolic quotients, e.g., $\mathcal{M}=\mathbb{H}^{3}/\Gamma$ where $\Gamma$ is a discrete subgroup of the isometry group $PSL(2,\mathbb{C})$ of $\Ht$, we can use the method of images \cite{Giombi:2008vd}
\beq 
K^{\mathbb{H}^{3}/\Gamma, (s)}_{ab}(x,y;t)=\sum_{\gamma\in\Gamma}K^{\mathbb{H}^{3}, (s)}_{ab}(x,\gamma y;t)\;,
\label{methimbasic}
\eeq
where $\gamma$ are the generators of $\Gamma$. Quotient spacetimes $\Ht/\Gamma$ are of interest, for example, when calculating 1-loop quantum corrections to holographic entanglement entropy \cite{Barrella:2013wja}.



The background $\mathcal{M}=\mathbb{H}^{3}/\Gamma$ is an example of a homogeneous space $\mathcal{M}=G/H$ \cite{David:2009xg}.
Not only are such spacetimes of physical interest, but their high degree of symmetry allows one to use group theoretic techniques to write down the eigenfunctions $\psi^{(s)}_{n,a}$ of the spin-$s$ Laplacian $\nabla^{2}_{(s)}$ in terms of matrix elements of representations of the symmetry group of $\mathcal{M}$. These group theoretic techniques were developed by Camporesi and Higuchi \cite{Higuchi:1986wu,Camporesi:1990wm,Camporesi:1992tm,Camporesi:1994ga,Camporesi:1995fb} and adapted to thermal AdS by \cite{David:2009xg,Gopakumar:2011qs}. We now summarize \cite{David:2009xg} and recast the findings in \cite{Keeler:2018lza} in the more illuminating language of group characters. 


The first task is to construct the heat kernel of a free field of spin-$s$ on Euclidean $\text{AdS}_{3}$. Euclidean $\text{AdS}_{3}$ is a homogeneous space $\Ht=SO(3,1)/SO(3)\simeq SL(2,\mathbb{C})/SU(2)$.  
Constructing wavefunctions  on $\Ht$ requires we define a section $\sigma$ in $SL(2,\mathbb{C})$ using representations of $SL(2,\mathbb{C})$. Using these representations the heat kernel on $\Ht$ can be written down explicitly \cite{David:2009xg}.


In this article we focus on thermal $\text{AdS}_{3}$
and the BTZ black hole, which both have quotient structure $\mathbb{H}^{3}/\Gamma$ with $\Gamma\simeq\mathbb{Z}$. Following (\ref{methimbasic}), the heat kernel on thermal $\text{AdS}_{3}$ ($\Ht/\mathbb{Z}$) can be calculated from the heat kernel on $\Ht$ via the method of images:
\beq K^{\Ht/\mathbb{Z},(s)}_{ab}(x,y;t)=\sum_{k\in\mathbb{Z}}K^{\Ht,(s)}_{ab}(x,\gamma^{k}(y);t)\;.\eeq
The trace of the coincident heat kernel $K(x,x;t)$ involves group integrals and traces over the appropriate representations of $SL(2,\mathbb{C})$. Upon integrating over the fundamental domain\footnote{The volume of the fundamental domain of $\mathbb{H}^3/\mathbb{Z}$ is infinite. In the second line of (\ref{fullhkspin}) this infinite volume term, corresponding to the $k=0$ piece, has been dropped. This is common practice, as it is removable by suitable regulatory methods. See for example \cite{Giombi:2008vd,David:2009xg}.} of $\Ht/\mathbb{Z}$, the integrated heat kernel for thermal $\text{AdS}_{3}$ for fields of spin $s$ is \cite{David:2009xg}
\beq
\begin{split}
 K^{\Ht/\mathbb{Z},(s)}(\tau,\bar{\tau};t)&=2\pi\tau_{2}\sum_{k\in\mathbb{Z}}\int^{\infty}_{0}d\lambda\chi_{\lambda,s}(e^{ik\pi \tau})e^{-t(\lambda^{2}+s+1)}\\
&\longrightarrow~\sum_{k=1}^{\infty}(-1)^{2ks}\frac{2\pi\tau_{2}(2-\delta_{s,0})}{2\sqrt{4\pi t}|\sin k\pi\tau|^{2}}\cos(2\pi sk\tau_{1})e^{-\frac{(2\pi k\tau_{2})^{2}}{4t}}e^{-(s+1)t}\;.\label{fullhkspin}
\end{split}
\eeq
Here $\chi_{\lambda,s}(e^{ik\pi\tau})$ is the Harish-Chandra \emph{group character} of the $SL(2,\mathbb{C})$ element $M=\text{diag}(e^{i\pi\tau},e^{-i\pi\tau})$,
\beq \chi_{\lambda,s}(e^{ik\pi\tau})=\frac{1}{2}\frac{\cos(2\pi s\tau_{1}-2\pi k\lambda\tau_{2})}{|\sin(k\pi\tau)|^{2}}\;,\label{character1}\eeq
and $\lambda$ are the weights of the associated Lie algebra. The term $(-1)^{2ks}$ is present to incorporate fermions with antiperiodic boundary conditions along the thermal circle. 

Equation (\ref{fullhkspin}) is for massless fields. The functional determinant of a kinetic operator of a massive spin-$s$ field on $\Ht/\mathbb{Z}$ is \cite{David:2009xg} given by 
\beq
\begin{split}
 -\log\text{det}(-\nabla^{2}_{(s)}+m^{2}_{s})&=\int^{\infty}_{0}\frac{dt}{t}K^{(s)}(\tau,\bar{\tau};t)e^{-m^{2}_{s}t}\\
&\rightarrow~\sum_{k=1}^{\infty}(-1)^{2ks}\frac{(2-\delta_{s,0})}{k|1-q^{k}|^{2}}(q^{ks}+\bar{q}^{ks})|q|^{k(\Delta_{s}-s)}\;,\label{funcdethk2}
\end{split}
\eeq
where $q\equiv e^{2\pi i\tau}, \bar{q}\equiv e^{-2\pi i\bar{\tau}}$ and we have defined
\beq \Delta_{s}\equiv1+\sqrt{(s+1)+m_{s}^{2}}\;.\label{confdim}\eeq
In the context of $\text{AdS}_{3}/\text{CFT}_{2}$, $\Delta_{s}$ represents the conformal dimension
of the $\text{CFT}_{2}$ operator dual to a spin-$s$ field propagating on $\text{AdS}_{3}$. 



\subsection{Quasinormal Mode Method}

We outline the quasinormal mode
method for computing 1-loop partition functions via a concrete example: a real massive scalar field $\phi$ in a rotating BTZ black hole \cite{Castro:2017mfj}. The quasinormal modes for higher spin fields in this background were computed in \cite{Birmingham:2001pj,Datta:2011za,Datta:2012gc}. 

We will make use of the BTZ line element written in global coordinates \cite{Carlip:1995qv}
\beq ds^{2}=d\xi^{2}-\sinh^{2}\xi dT^{2}+\cosh^{2}\xi d\Phi^{2}\;,\label{regcoord}\eeq
where  $\Phi$ and $\xi$ behave as angular and radial coordinates, respectively. The behavior of a real massive scalar field near the horizon $r\sim r_{+}$ is 
\beq \phi(\xi,T_{E},\Phi)\sim\xi^{\pm ik_{T}}e^{-k_{T}T_{E}}e^{-ik_{\Phi}\Phi}\;.\label{scalarfield}\eeq
Here $k_{T}=\frac{\omega r_{+}-ik|r_{-}|}{r^{2}_{+}+|r_{-}|^{2}}$ is the frequency conjugate to the Euclidean time coordinate $T_{E}$ 
and $k_{\Phi}$ is conjugate to the coordinate $\Phi$, while $k\in\mathbb{Z}$ is the angular momentum quantum number. 

Periodicity of $\phi$ in the $T_E$ direction requires that $k_{T}=in$ for $n\in\mathbb{Z}$. Solutions (\ref{scalarfield}) will only occur at specific quantized values $\omega_n$ --- the Matsubara frequencies:
%
%
%
\beq 
-ik_{T}=n\Rightarrow \frac{\omega_{n}}{2\pi}=2i\frac{T_{L}T_{R}}{T_{L}+T_{R}}n+\frac{T_{R}-T_{L}}{T_{L}+T_{R}}\frac{k}{2\pi}\;,\label{Matfreqrot}
\eeq
where $T_{L,R}=\frac{1}{2\pi}(r_{+}\mp r_{-})$.  The $n>0$ quasinormal modes are ingoing while the $n<0$ antiquasinormal modes are outgoing. Following \cite{Castro:2017mfj}, we may rewrite (\ref{Matfreqrot})
\beq
\frac{\omega_{n}}{2\pi}=\frac{in}{2}(T_{R}+T_{L})-\frac{(T_{R}-T_{L})}{2}k_{\Phi}(n,k)\;,\quad k_{\Phi}(n,k)=\frac{(T_{R}-T_{L})}{(T_{R}+T_{L})}in-\frac{k}{\pi(T_{R}+T_{L})}\;,\label{introducekphi}
\eeq

The ingoing quasinormal mode frequencies are \cite{Birmingham:2001pj}: 
\beq
\omega_{\ast}=-k-2\pi iT_{R}(2p+\Delta)\;,\quad \omega_{\ast}=k-2\pi i T_{L}(2p+\Delta)\;, \label{ingoingQNMs}
\eeq
while the outgoing antiquasinormal frequencies are 
\beq \omega_{\ast}=-k+2\pi iT_{R}(2p+\Delta)\;,\quad \omega_{\ast}=k+2\pi i T_{L}(2p+\Delta)\;.\label{outgoingQNMs}\eeq
Here $p \in \mathbb{N}$ is the radial quantum number and $\Delta$ is the conformal dimension.

Assuming $Z^{(1)}(\Delta)$ is meromorphic, the quasinormal mode method \cite{Denef:2009yy,Denef:2009kn} allows us to exploit the Weierstrass factorization theorem
, permitting us to write a meromorphic function $Z^{(1)}(\Delta)$ as a product of its zeros and poles, up to an entire function $e^{\text{Poly}(\Delta)}$:
\beq 
Z^{(1)}(\Delta)=e^{\text{Poly}(\Delta)}\frac{\prod_{\Delta_{0}}(\Delta-\Delta_{0})^{d_{0}}}{\prod_{\Delta_{p}}(\Delta-\Delta_{p})^{d_{p}}}\;.
\label{Weierstrassfact}
\eeq
Here $d_{0}$ and $d_{p}$ are the degeneracies of the zeros $\Delta_{0}$ and poles $\Delta_{p}$, respectively, and $\text{Poly}(\Delta)$ is a polynomial with only positive powers of $\Delta$.  Since $Z^{(1)}_{\text{scalar}}\propto(\text{det}\nabla^2)^{-1/2}$, it has no zeros but will have a pole whenever $\nabla^2$ has a zero mode.  These zero modes occur when $\Delta$ is tuned such that the Klein-Gordon equation,
\beq
\left(-\nabla^{2}+\Delta_{n,\ast}(\Delta_{n,\ast}-2)\right)\phi=0,\;
\label{kgeqn}
\eeq
has a smooth, single-valued solution for $\phi$ in Euclidean signature which obeys the asymptotic boundary conditions. We denote such solutions  by $\phi_{\ast,n}$, where $n$ labels the mode number in the Euclidean time direction and $\ast$ represents the $p$ and $\ell$ quantum numbers. The associated $\Delta$ for which $\phi_{\ast,n}$ solve the Klein-Gordon equation are likewise denoted as $\Delta_{\ast,n}$. Therefore, poles in $Z^{(1)}(\Delta)$ occur whenever $\Delta=\Delta_{\ast,n}$. 

Furthermore, when $\Delta$ is tuned to $\Delta_{\ast,n}$, the quasinormal modes $\omega_{\ast}(\Delta_{\ast,n})$  (\ref{ingoingQNMs}) and antiquasinormal modes (\ref{outgoingQNMs})  align with the Matsubara frequencies $\omega_{n}$ (\ref{Matfreqrot}):
\beq
\omega_{\ast}(\Delta_{\ast,n})=\omega_{n}\;\label{omegaast=omegan}.
\eeq
Thus, tuning the conformal dimension $\Delta$ such that $\omega_{\ast}(\Delta)=\omega_{n}$ provides us with the location of the poles of $Z^{(1)}(\Delta)$. Using this insight, the 1-loop determinant of a real scalar field on a rotating BTZ black hole background was calculated in \cite{Castro:2017mfj}:
\beq 
\begin{split}
\left(\frac{e^{\text{Poly}(\Delta)}}{Z^{(1)}}\right)^{2}&=\prod_{n>0,p\geq0,k}(\omega_{n}+k+2\pi iT_{R}(2p+\Delta))(\omega_{n}-k+2\pi iT_{L}(2p+\Delta))\\
&\prod_{n<0,p\geq0,k}(\omega_{n}+k-2\pi iT_{R}(2p+\Delta))(\omega_{n}-k-2\pi iT_{L}(2p+\Delta))\\
&\prod_{p\geq0,k}(\omega_{0}+k+2\pi iT_{R}(2p+\Delta))(\omega_{0}-k+2\pi iT_{L}(2p+\Delta))\;.
\end{split}
\label{1loopscalarQNM}\eeq
After absorbing a factor into Poly$(\Delta)$, (\ref{1loopscalarQNM}) can be recast as
\beq Z^{(1)}=e^{\text{Poly}(\Delta)}\prod_{\tilde{\ell},\tilde{\ell}'=0}^{\infty}\frac{1}{(1-q^{\tilde{\ell}+\Delta/2}\bar{q}^{\tilde{\ell}'+\Delta/2})}\;,\label{1loopscalarQNM2}\eeq
where $\tilde{\ell}$ and $\tilde{\ell}'$ are combinations of quantum numbers $p$ and $n$, $q=e^{-2\pi(2\pi T_{L})}$, and $\bar{q}=e^{-2\pi(2\pi T_{R})}$.

The 1-loop partition function for the arbitrary spin-$s$ case on a non-rotating BTZ was worked out in \cite{Datta:2011za}. Following the methods applied in \cite{Castro:2017mfj}, it is straightforward to generalize (\ref{1loopscalarQNM2}) to arbitrary spin-$s$ fields on a rotating BTZ background
\beq Z^{(1)}_{(s)}=e^{\text{Poly}(\Delta_{s})}\prod_{\tilde{\ell},\tilde{\ell}'=0}^{\infty}\frac{1}{\left(1-q^{\tilde{\ell}+\frac{\Delta_{s}}{2}-\frac{s}{2}}\bar{q}^{\tilde{\ell}'+\frac{\Delta_{s}}{2}+\frac{s}{2}}\right)\left(1-q^{\tilde{\ell}+\frac{\Delta_{s}}{2}+\frac{s}{2}}\bar{q}^{\tilde{\ell}'+\frac{\Delta_{s}}{2}-\frac{s}{2}}\right)}\;.\label{Z1loopQNMgens}\eeq
A quick way to derive this expression is to use (\ref{1loopscalarQNM2}) and shift the conformal dimension. For $m_{s}>0$, replace the scalar conformal dimension $\Delta\to\Delta_{s}+s\frac{T_{R}-T_{L}}{T_{R}+T_{L}}$, giving $Z^{(1)}_{(s),m_{s}>0}$, while for $m_{s}<0$ replace $\Delta\to\Delta_{s}-s\frac{T_{R}-T_{L}}{T_{R}+T_{L}}$, giving $Z^{(1)}_{(s),m_{s}<0}$. The total expression of the 1-loop partition function is $Z^{(1)}_{(s)}=Z^{(1)}_{(s),m_{s}>0}Z^{(1)}_{(s),m_{s}<0}$. We will show how the ``effective" conformal dimensions $\Delta_{s}\pm s\frac{T_{R}-T_{L}}{T_{R}+T_{L}}$ also arise from the zeros of the Selberg zeta function.


\subsection{Comparing the two Methods}
\indent

Let us now explicitly compare the heat kernel and quasinormal methods. A comparison between these two methods for arbitrary spin fields on a non-rotating BTZ black hole was previously shown in \cite{Datta:2011za,Datta:2012gc}. Here we compare the two methods for spin-$s$ bosons on a rotating BTZ black hole. 

The logarithm of (\ref{Z1loopQNMgens}) leads to
\beq
\begin{split}
\log Z^{(1)}_{(s)}-\text{Poly}(\Delta_{s})&=-\sum_{\tilde{\ell},\tilde{\ell}'=0}^{\infty}\left[\log\left(1-q^{\tilde{\ell}+\frac{\Delta_{s}}{2}-\frac{s}{2}}\bar{q}^{\tilde{\ell}'+\frac{\Delta_{s}}{2}+\frac{s}{2}}\right)+\log\left(1-q^{\tilde{\ell}+\frac{\Delta_{s}}{2}+\frac{s}{2}}\bar{q}^{\tilde{\ell}'+\frac{\Delta_{s}}{2}-\frac{s}{2}}\right)\right]\\
&=\sum_{k=1}^{\infty}\sum_{\tilde{\ell},\tilde{\ell}'=0}^{\infty}\frac{1}{k}\left[\left(q^{\tilde{\ell}+\frac{\Delta_{s}}{2}-\frac{s}{2}}\bar{q}^{\tilde{\ell}'+\frac{\Delta_{s}}{2}+\frac{s}{2}}\right)^{k}+\left(q^{\tilde{\ell}+\frac{\Delta_{s}}{2}+\frac{s}{2}}\bar{q}^{\tilde{\ell}'+\frac{\Delta_{s}}{2}-\frac{s}{2}}\right)^{k}\right]\\
&=\sum_{k=1}^{\infty}\frac{(q\bar{q})^{k\left(\frac{\Delta_{s}}{2}-\frac{s}{2}\right)}}{k}\sum_{\tilde{\ell},\tilde{\ell}'=0}^{\infty}q^{k\tilde{\ell}}\bar{q}^{k\tilde{\ell}'}(q^{sk}+\bar{q}^{sk})\\
&=\sum_{k=1}^{\infty}\frac{|q|^{k\left(\Delta_{s}-s\right)}}{k|1-q^{k}|^{2}}(q^{ks}+\bar{q}^{ks})\;,
\end{split}
\label{logZ1loopqnm}\eeq
where in the second line we used the series expansion for $\log(1-x)$. The above expression (\ref{logZ1loopqnm}) should be compared to the expression found using the heat kernel method (\ref{funcdethk2})\footnote{From first appearances, the two expressions are not identical. The object computed using the heat kernel method has a factor of $(-1)^{2ks}$ and $(2-\delta_{s,0})$. The first of these factors can be incorporated in a straightforward way by allowing for both periodic or antiperiodic boundary conditions of spinor fields in \cite{Datta:2012gc}. The factor $(2-\delta_{s,0})$ can also be put in by hand to account for the scalar contribution to $\log(Z^{(1)}_{(s)})$. We also point out that (\ref{logZ1loopqnm}) includes a $\text{Poly}(\Delta)$ contribution, which is proportional to the volume of $\Ht/\mathbb{Z}$, and can be removed via a proper regularization scheme \cite{Vassilevich:2003xt}, much like how the divergent volume contribution present in (\ref{funcdethk2}) is dealt with.}.

We see then that the sum over $k$, here introduced when we took the logarithm of the 1-loop partition function, is identified with the sum over images in the heat kernel approach, while $\tilde{\ell}$ and $\tilde{\ell}'$ --- a combination of the thermal number $n$ and radial quantum number $p$ --- build the integration measure appearing in the trace of the coincident heat kernel, agreeing with the observation from \cite{Keeler:2018lza}. More precisely, we find that the combined sum over the quantum number $p$ and thermal integer $n$ builds the group character (\ref{character1}) of $SL(2,\mathbb{C})$ coming from the structure of the underlying homogeneous space $\Ht\simeq SL(2,\mathbb{C})/SU(2)$.


The fact that quasinormal modes encode information about the group structure of the underlying homogeneous space is not entirely surprising. This is because the quasinormal modes Wick rotate into Euclidean zero modes and, with $\Delta_{n,\ast}$ restricted to the negative integers, are unitary representations with respect to the norm on $\Ht$. These unitary representations, moreover, are equivalent to discrete series representations on $SL(2,\mathbb{C})$ \cite{Keeler:2014hba}. Above we have explicitly shown that the quantum numbers appearing in the quasinormal frequencies carry information about the Harish-Chandra character of an element of $SL(2,\mathbb{C})$.

\section{Quasinormal Modes from the Zeros of the Selberg Zeta Function}\label{sec:connectingonH3}

We now provide a more formal connection between the quasinormal mode and heat kernel methods for computing 1-loop determinants with respect to arbitrary spin-$s$ fields on quotient spaces $\Ht/\mathbb{Z}$, extending the discussion in \cite{Keeler:2018lza} to general spin. In \cite{Keeler:2018lza} we connected the heat kernel and quasinormal mode methods via the Selberg zeta function $Z_{\Gamma}$, a zeta function associated with a hyperbolic quotient, e.g., $\mathbb{H}^3/\Gamma$, with $\Gamma$ as a discrete subgroup of $SL(2,\mathbb{C})$. Explicitly, the Selberg zeta function of the (Euclidean) BTZ black hole, i.e., the quotient space $\Ht/\mathbb{Z}$, was derived in \cite{Perry03-1}:
\beq Z_{\mathbb{Z}}(z)=\prod_{k_{1},k_{2}=0}^{\infty}\left[1-e^{2ibk_{1}}e^{-2ibk_{2}}e^{-2a(k_{1}+k_{2}+z)}\right]\;.\label{Eulerzeta2}\eeq
Here parameters $a$ and $b$ are related to the action of the generator $\gamma$ of $\mathbb{Z}$, where $a$ is related to 
the length of closed, primitive geodesics on $\Ht/\mathbb{Z}$. In practice we may relate $a$ and $b$ to the geometry of the BTZ black hole\footnote{Recall that $\Ht/\mathbb{Z}$ is a solid torus whose $T^{2}$ boundary has modulus $\tau$, arising from the periodic identification of the Euclidean time coordinate $t_{E}$ and angular coordinate $\phi$, $(t_{E},\phi)\sim(t_{E}+\beta,\phi+\theta)$. $\tau$ is a combination of the inverse temperature $\beta$ and angular potential $\theta$ of $\Ht/\mathbb{Z}$, $2\pi\tau=(\theta+i\beta)$.}, specifically to the modular parameter $\tau=\tau_{1}+i\tau_{2}$ characterizing the $T^{2}$ boundary of $\Ht/\mathbb{Z}$, 
\beq a=\pi \tau_{2}=\pi r_{+}=\pi^{2}(T_{R}+T_{L})\;,\quad b=-\pi\tau_{1}=\pi|r_{-}|=i\pi^{2}(T_{R}-T_{L})\;.\label{abparams}\eeq



The Selberg zeta function (\ref{Eulerzeta2}) converges for all $z$ \cite{Perry03-1}, and is an entire function with zeros occurring whenever the argument of the exponential equals $2\pi i\ell$ for $\ell\in\mathbb{Z}$:
\beq z_{\ell,k_{1},k_{2}}^{\ast}=-(k_{1}+k_{2})+\frac{ib}{a}(k_{1}-k_{2})-\frac{i\pi \ell}{a}\;. \label{PSzetazeros}\eeq
In \cite{Keeler:2018lza} we demonstrated that identifying the zeros $z^{\ast}$ of the Selberg zeta function with the conformal dimension $\Delta$ of the field is equivalent to identifying the quasinormal mode frequencies with the Matsubara frequencies (\ref{equivcond}), that is, to setting $\omega_{\ast}=\omega_{n}(T)$. Here we extend this result to fields of arbitrary spin.

As was reported in \cite{Keeler:2018lza,DHoker:1985een}, we may write the 1-loop determinant of a scalar field on $\Ht/\mathbb{Z}$  directly in terms of the Selberg zeta function
\beq -\log\text{det}(-\nabla^{2}+m^{2})=-2\log Z_{\Gamma}(\Delta)\;,\label{funcdetselbs0}\eeq
where we have again neglected the infinite volume contribution. Let us now extend (\ref{funcdetselbs0}) to include arbitrary spin-$s$ fields. We begin with bosonic fields.

\section*{Bosons}
\indent

For bosonic fields with $s\neq0$ we start from the 1-loop determinant, e.g., (\ref{funcdethk2}), and recast it in terms of a product of Selberg zeta functions \cite{l2015zeta}
\beq \log\text{det}(-\nabla^{2}_{(s)}+m^{2}_{s})_{s\in\mathbb{Z}^{+}}=\log \left[Z_{\Gamma}\left(\Delta_{s}+\frac{isb}{a}\right)\cdot Z_{\Gamma}\left(\Delta_{s}-\frac{isb}{a}\right)\right]\;.\label{funcdetselbgensbos}\eeq
Here $\Delta_{s}$ is the conformal dimension of the $\text{CFT}_{2}$ operator dual to a spin-$s$ field living in $\text{AdS}_{3}$, (\ref{confdim}). We can apply equation (\ref{funcdetselbgensbos}) and write the 1-loop partition function for various types of fields on a rotating BTZ background in terms of Selberg zeta functions, e.g., the massless graviton, as in \cite{l2015zeta}. The arguments $\Delta_{s}\pm \frac{isb}{a}$ arise because we can write the spin $s$ partition function for bosons in a rotating BTZ background as
\begin{equation}
\begin{split}
 Z^{(1)}_s(\Delta_s)&=Z^{(1)}_{m_s<0}(\Delta_s)Z_{m_s>0}^{(1)}(\Delta_s)\\&=Z^{(1)}_{\text{scalar}}\Big(\Delta_s+\frac{isb}{a}\Big)Z_{\text{scalar}}^{(1)}\Big(\Delta_s-\frac{isb}{a}\Big).
\end{split}
\end{equation}

The quasinormal mode method of computing 1-loop partition functions involves first solving for the quasinormal frequencies $\omega_{\ast}$ and then identifying the Euclidean zero modes of the corresponding wave equation. Regularity of the Euclidean zero modes imposes the condition that the poles of the 1-loop partition function occur whenever $\omega_{\ast}=\omega_{n}$. As detailed in Appendix B of \cite{Castro:2017mfj}, there is an additional restriction on specific states, arising from the square-integrability of the Euclidean solutions near the tip of the Euclidean cigar. For example, for a massive spin-2 excitation $h_{\mu\nu}$, Euclidean solutions $h^{(\lambda)}_{\mu\nu}$ are required to satisfy the integrability condition \cite{Camporesi:1994ga,Castro:2017mfj}
\beq \int d^{3}x\sqrt{g}g^{\mu\nu}g^{\rho\sigma}h^{(\lambda)}_{\mu\rho}(x)h^{(\lambda')\ast}_{\nu\sigma}(x)=\delta(\lambda-\lambda')\;,\label{sqint}\eeq
where $\lambda$ is an eigenvalue. Notably, to avoid non-integrable solutions at the singularity $\xi=0$ in (\ref{regcoord}), the range over the thermal integer $n$ must be further restricted for Euclidean solutions with certain low-lying values of the radial quantum number $p$. Such states are discarded, and the remaining solutions are referred to as ``good" Euclidean zero modes \cite{Castro:2017mfj}. 

The  matching condition $\omega_{\ast}=\omega_{n}$ for the ``good" Euclidean modes was worked out for the massive spin-2 and spin-1 bosons in \cite{Castro:2017mfj}. The equivalent conditions for arbitrary spin-$s$ fields are displayed in Table \ref{matching}. For convenience we restate the quantity $k_{\Phi}(n,k)$ (\ref{introducekphi}) in terms of parameters $a$ and $b$:
\beq ik_{\Phi}(n,k)=\frac{ib}{a}n-\frac{i\pi}{a}k\;.\label{kphiab}\eeq

It is instructive to recast the condition $\omega_{\ast}(\Delta_s)=\omega_{n}$ for a spin-s field (Table \ref{matching}) in terms of the equivalent condition for scalar fields, i.e. $\omega_{\ast}^{\text{scalar}}(\Delta)=\omega_{n}$. To do this we summarize the four conditions in Table \ref{matching} into a single expression:
\beq 2p+\Delta_{s}+|n-\sigma_{s}s|+i\sigma_{s}\sigma_{m}k_{\Phi}(n,k)=0\;.\label{tab1sum}\eeq
Here $\sigma_{s}=\pm$ corresponding to the sign of the shift to $n$ by $s$, and $\sigma_{m}=\pm$ corresponds to the sign of the field mass $m_{s}$, e.g., $\sigma_{m}=(-1)$ when $m_{s}>0$.

Relabeling the integer $n$ by $n+\sigma_{s}s$ we find that the spin-dependent terms in (\ref{tab1sum}) are
\beq \Delta_{s}+i\sigma_{m}\frac{sb}{a}\;.\eeq
Consequently, (\ref{tab1sum}) becomes the condition
\beq \omega_{\ast}^{\text{scalar}}\left(\Delta_{s}+i\sigma_{m}\frac{sb}{a}\right)=\omega_{n}\;.\eeq
For example, when $m_{s}>0$, and $\sigma_{s}=(-1)$, we have that the top left condition in Table \ref{matching} becomes $\omega_{\ast}^{\text{scalar}}\left(\Delta_{s}-i\frac{sb}{a}\right)=\omega_{n}$. Thus we see the arguments $\Delta_{s}\pm\frac{isb}{a}$ of the Selberg zeta functions (\ref{funcdetselbgensbos}) arise as a consequence of discarding the unphysical non-square integrable Euclidean zero modes, and rewriting the conditions in Table \ref{matching} in the same manner as for scalar fields.

\vspace{1mm}

{\renewcommand{\arraystretch}{1.5}
\begin{table}
\begin{center}
\begin{tabular}{|c|c|}
\hline 
 $m_{s}>0$ & $m_{s}<0$ \\ \cline{1-2} 
 $2p+\Delta_{s}+|n+s|+ik_{\Phi}(n,k)=0$ & $2p+\Delta_{s}+|n-s|+ik_{\Phi}(n,k)=0$ \\
 $2p+\Delta_{s}+|n-s|-ik_{\Phi}(n,k)=0$  & $2p+\Delta_{s}+|n+s|-ik_{\Phi}(n,k)=0$\\ 
  \hline
\end{tabular}
\caption{The conditions satisfied by square-integrable Euclidean solutions. Recall that for bosonic fields $n,k\in\mathbb{Z}$. These equations reproduce the condition $\omega_{n}=\omega_{\ast}$ \cite{Castro:2017mfj}.}
\label{matching}
\end{center}
\end{table}

Let us now elucidate the connection between the zeros of the Selberg zeta function and the quasinormal frequencies, generalizing \cite{Keeler:2018lza}. Setting the arguments $\Delta_{s}+i\sigma_{m}\frac{isb}{a}$ of the 1-loop partition function (\ref{funcdetselbgensbos}), to the zeros $z^{\ast}$ (\ref{PSzetazeros}) leads to 
\beq k_{1}+k_{2}+\Delta_{s}-\left[\frac{ib}{a}(k_{1}-k_{2}-\sigma_{m}s)-\frac{i\pi\ell}{a}\right]=0\;.\label{deltamsg0}\eeq
Notice that when we relabel the positive integers $k_{1},k_{2}$ as
\beq k_{1}+k_{2}=2p+|n-\sigma_{s}s|\;,\quad k_{1}-k_{2}=-\sigma_{s}\sigma_{m}n+\sigma_{m}s\;,\quad \ell=-\sigma_{m}\sigma_{s}k\;,\label{relabelsum}\eeq
we have that (\ref{deltamsg0}) is equivalently written as (\ref{tab1sum}), the collection of conditions $\omega_{n}=\omega_{\ast}$ displayed in Table \ref{matching}.
For example, when $\sigma_{m}=(-1)$ and $\sigma_{s}=(-1)$, we have the relabeling
\beq k_{1}+k_{2}=2p+|n+s|\;,\quad k_{1}-k_{2}=-(n+s)\;,\quad \ell=-k\;,\eeq
leading to 
\beq 2p+|n+s|+ik_{\Phi}(n,\ell)=0\;,\label{BLmsg0}\eeq
the top left condition in Table \ref{matching}. 

We see that there are only four possible\footnote{We could have tried choosing $\sigma_{m}=(+1)$ but with the relabeling $k_{1}+k_{2}=2p+|n+s|$, $k_{1}-k_{2}=-n+s$. Doing so we would also arrive to (\ref{BLmsg0}). Note, however, these redefinitions of $k_{1}$ and $k_{2}$ result in the radial quantum number satisfying $p<0$ for non-zero $s$, which is inconsistent with the fact that $p$ must be a positive integer. Therefore Table \ref{relabelk1k2} gives the only consistent relabelings that reproduce the conditions in Table \ref{matching}.} choices of relabeling $k_{1},k_{2}$. The four options come from the two signs, $\sigma_{m}$ and $\sigma_{s}$. These four signs fix four redefinitions of $k_{1},k_{2}$, which we list in Table \ref{relabelk1k2}. 

We emphasize that our relabeling of integers $k_{1},k_{2}$ is well-motivated from spectral theory on hyperbolic quotients  \cite{Perry03-1}. Specifically, the integers of $k_{1}$ and $k_{2}$ are repackaged into new integers $p=0,1,2,...$ and $n\in\mathbb{Z}$ 
\beq
\begin{split}
&n\geq 0: \qquad k_{1}+k_{2}=2p+n \qquad  k_{1}-k_{2}=\mp n\;,\\
&n<0: \qquad k_{1}+k_{2}=2p-n \qquad  k_{1}-k_{2}=\mp n\;,
\end{split}
\label{redefs}
\eeq
such that the zeros of the Selberg zeta function are identified with the poles of the so-called ``scattering" operator $\Delta_{\Gamma}$\footnote{As detailed in \cite{Perry03-1}, the scattering operator $\Delta_{\Gamma}$ is the positive Laplacian acting on the Hilbert space $\mathcal{H}=L^{2}((0,\infty)\times(0,2a)\times(0,2\pi)\sinh\xi\cosh\xi d\xi d\Phi dT_{E})$.}. Our relabeling in Table \ref{relabelk1k2} can be understood as a spin-$s$ generalization of the identifications made in \cite{Perry03-1}, where we have also removed non-square-integrable Euclidean zero modes.


\vspace{1mm}

{\renewcommand{\arraystretch}{1.25}
\begin{table}
\begin{center}
\begin{tabular}{|c|c|c|}
\hline 
 {} & $\Delta_{s}-\frac{isb}{a}\;(m_{s}>0)$ & $\Delta_{s}+\frac{isb}{a}\;(m_{s}<0)$ \\ \cline{1-3} 
 $\ell=-k$ & $k_{1}+k_{2}=2p+|n+s|$ & $k_{1}+k_{2}=2p+|n-s|$ \\
 {} &  $k_{1}-k_{2}=-(n+s)$ &  $k_{1}-k_{2}=-(n-s)$ \\ \cline{1-3}
$\ell=k$ & $k_{1}+k_{2}=2p+|n-s|$ & $k_{1}+k_{2}=2p+|n+s|$\\ 
{} & $k_{1}-k_{2}=n-s$ & $k_{1}-k_{2}=n+s$ \\
  \hline
\end{tabular}
\caption{Relabeling positive integers $k_{1}$ and $k_{2}$ from setting the arguments $\Delta_{s}\pm\frac{isb}{a}$ to the Selberg zeros. The four possible choices of relabeling of $k_{1}$ and $k_{2}$ come from the four combinations of the signs $\sigma_{s}$ and $\sigma_{m}$, and is neatly summarized by (\ref{relabelsum}). These redefinitions of $k_{1}$ and $k_{2}$ produce the $\omega_{n}=\omega_{\ast}$ conditions displayed in Table \ref{matching}.}
\label{relabelk1k2}
\end{center}
\end{table}

\section*{Fermions}
\indent

The 1-loop determinant for fermionic fields with anti-periodic boundary conditions in the Euclidean time direction\footnote{We may also consider fermions with periodic boundary conditions along the thermal circle, where we drop the $(-1)^{k}$ appearing in the sum over images of the functional determinant, thereby removing the $\pm\frac{i\pi}{2a}$ from the argument of the Selberg zeta function in (\ref{funcdetselbgensferm}). The resulting 1-loop partition function then requires the introduction of the $(-1)^{F}$ operator, where $F$ is the fermion number operator \cite{David:2009xg,DiFrancesco:1997nk}. Even in the periodic case, however, since $k_{1},k_{2}$ are positive integers and $s$ is a half integer, we require that $n$ be a half integer.} takes the form
\beq \log\text{det}(-\nabla^{2}_{(s)}+m^{2}_{s})_{s\in\mathbb{Z}^{+}_{\frac{1}{2}}}=\log \left[Z_{\Gamma}\left(\Delta_{s}+\frac{isb}{a}+\frac{i\pi }{2a}\right)\cdot Z_{\Gamma}\left(\Delta_{s}-\frac{isb}{a}+\frac{i\pi }{2a}\right)\right]\;,\label{funcdetselbgensferm}
\eeq
where we used that for $s\in\mathbb{Z}^{+}+\frac{1}{2}$, $(-1)^{2ks}=(-1)^{k}$. The authors of \cite{l2015zeta} used this expression to write the 1-loop partition function for the Majorana gravitino in $\mathcal{N}=1$ supergravity.

Consider the arguments $\Delta_s+i\sigma_{m}\frac{sb}{a}+\frac{i\pi}{2a}$. In the bosonic case we saw that this argument arose from the removal of non-square-integrable zero modes, Table \ref{matching}. We now ask how the conditions in Table \ref{matching} change for fermionic fields. Anti-periodic boundary conditions in the Euclidean time direction means that we are to shift $n\rightarrow n+1/2$ (e.g.,\cite{Denef:2009kn}). A priori, we still have a choice to impose periodic (Ramond) or anti-periodic (Neveu-Schwarz) boundary conditions along the $k$ cycle of the torus \cite{DiFrancesco:1997nk}. We will see now that anti-periodic boundary conditions along the $k$-cycle reproduce the argument $\Delta_{s}+i\sigma_{m}\frac{sb}{a}+\frac{i\pi}{2a}$, so we will select these boundary conditions. 

Take the conditions in Table \ref{matching}, summarized by (\ref{tab1sum}), now with $n\to n+1/2$, and $k\to k+1/2$. Again, as in the bosonic case, shifting $n\to n+\sigma_{s}s$  gives us the condition $\omega_{\ast}^{\text{scalar}}(\Delta_{s}+i\sigma_{m}\frac{sb}{a}+\frac{i\pi}{2a})=\omega_{n}$. Setting the arguments $\Delta_s+i\sigma_{m}\frac{sb}{a}+\frac{i\pi}{2a}$ appearing in the expression for the 1-loop partition determinant for fermions (\ref{funcdetselbgensferm}) to the zeros of the Selberg zeta function (\ref{PSzetazeros}) yields
\beq \Delta_{s}+k_{1}+k_{2}-\left[\frac{ib}{a}(k_{1}-k_{2}-\sigma_{m}s)-\frac{i\pi}{a}\left(\ell+\frac{1}{2}\right)\right]=0\;.\label{deltamsg0ferm}\eeq 
The relabeling of integers $k_{1},k_{2}$ for fermions now generalizes (\ref{relabelsum}) in a straightforward way:
\beq k_{1}+k_{2}=2p+\biggr|n+\frac{1}{2}-\sigma_{s}s\biggr|\;,\quad k_{1}-k_{2}=-\sigma_{s}\sigma_{m}\left(n+\frac{1}{2}\right)+\sigma_{m}s\;,\quad \ell+\frac{1}{2}=-\sigma_{s}\sigma_{m}k\;.\label{relabelfermsum}\eeq
Substituting this relabeling into (\ref{deltamsg0ferm}) gives us 
\beq 2p+\Delta_{s}+\biggr|n+\frac{1}{2}-\sigma_{s}s\biggr|+i\sigma_{s}\sigma_{m}k_{\Phi}(n+1/2,k+1/2)=0\;.\eeq
This expression is the fermionic equivalent of the $\omega_{n}=\omega_{\ast}$ conditions displayed for bosons in Table \ref{matching}.




In summary, the zeros of the Selberg zeta function --- an object built entirely out of the group structure of the underlying spacetime --- encode the quasinormal frequencies of fields which propagate on this spacetime. In particular, setting the arguments of the Selberg zeta function expressions in (\ref{funcdetselbgensbos}) or  (\ref{funcdetselbgensferm}) equal to the zeros of the Selberg zeta function leads to the condition that the Matsubara frequencies match the quasinormal frequencies:
\beq \Delta_{s}\pm\frac{isb}{a}=z^{\ast}\Longleftrightarrow \omega_{n}=\omega_{\ast}(\Delta_{s})\;.\eeq



\section{Discussion}
\label{disc}

We have extended the connection between the heat kernel and quasinormal mode methods for computing 1-loop functional determinants on rotating BTZ to arbitrary spin-$s$ fields. Upon writing the 1-loop partition function as a product of Selberg zeta functions \cite{l2015zeta}, we showed that by setting their arguments $\Delta_{s}\pm\frac{isb}{a}$ to the zeros of the Selberg zeta function we recover the Matsubara matching condition $\omega_{\ast}=\omega_{n}$. The heat kernel and quasinormal mode methods are thus connected by the Selberg zeta function.


An interesting consequence of this work is that we obtained a generalization of the relabeling of integers $k_{1},k_{2}$ given in \cite{Perry03-1} for the case of arbitrary spin fields. While \cite{Perry03-1} derived these relabelings from scattering theory on $\Ht/\mathbb{Z}$, our derivation instead arises via the connection between the heat kernel and quasinormal modes. In the case of fermions, relating the Selberg zeta function (\ref{funcdetselbgensferm}) to the 1-loop partition function appears to pick out a natural set of boundary conditions, i.e., (NS,NS). 


There are possible extensions to this work. First we can extend our analysis to higher dimensional thermal $\text{AdS}$ spacetimes. Specifically, we can consider the Selberg zeta function on $\mathbb{H}^{2n+1}/\mathbb{Z}$, and, together with the Matsubara frequencies, show that the Selberg zeros encode the \emph{normal} frequencies of spin-$s$ fields living on thermal $\text{AdS}_{2n+1}$. Once these frequencies are known, we can then construct the corresponding heat kernels on thermal $\text{AdS}_{2n+1}$. This extension --- which is expected to appear soon --- may lead to a deeper understanding of Vasiliev theories \cite{Vasiliev:2003ev}, consistent classical descriptions of bulk $\text{AdS}_{d+1}$ in the AdS/CFT correspondence.


Another extension would be to adapt our formalism to non-hyperbolic spacetimes that possess sufficient symmetry. For example, we could consider the sphere $S^{3}$ and the quotients $S^{3}/\Gamma$, e.g., Lens spaces $S^{3}/\mathbb{Z}_{p}$. In fact, it is straightforward to extend our analysis from $\text{AdS}_{3}$ to $S^{3}$ via a formal Wick rotation, where the normal modes of thermal AdS become the quasinormal modes of Euclidean de Sitter space. Understanding the relationship between the heat kernel and quasinormal mode methods of computing 1-loop determinants on $S^{3}$ and $S^{3}/\Gamma$ may lead to deeper insights into the de Sitter Farey tail and de Sitter quantum gravity \cite{Castro:2011xb}. Currently this work is underway.

Finally, another forthcoming work focuses on the study of quasinormal modes on product spaces of the form $S^N\times S^M$. Product spaces such as these naturally possess a quotient structure. Although Selberg zeta functions only exist for hyperbolic (rather than spherical) quotients, once the corresponding connection between the quasinormal mode and heat kernel methods for spheres is established, building the connection for product spaces is a natural extension.

\section*{Acknowledgements}

We would like to David McGady and Alankrita Priya for helpful discussions. The work of CK and VM is supported by the U.S. Department of Energy under grant number DE-SC0018330.


\appendix


\bibliography{QNMgensbib}

\end{document}